\documentstyle[12pt,epsfig,rotating]{article}
\def\bsls35{\baselineskip 0.35in}

\def\bd{\begin{document}} \def\ed{\end{document}}
\def\bmp{\begin{minipage}} \def\emp{\end{minipage}}
\def\bcc{\begin{center}} \def\ecc{\end{center}}     \def\npg{\newpage}
\def\beq{\begin{equation}} \def\eeq{\end{equation}} \def\hph{\hphantom}
\def\be{\begin{equation}} \def\ee{\end{equation}} \def\r#1{$^{[#1]}$}
\def\n{\noindent} \def\ni{\noindent} \def\pa{\parindent}
\def\hs{\hskip} \def\vs{\vskip} \def\hf{\hfill} \def\ej{\vfill\eject}
\def\cl{\centerline} \def\ob{\obeylines}  \def\ls{\leftskip}
\def\underbar#1{$\setbox0=\hbox{#1} \dp0=1.5pt \mathsurround=0pt
   \underline{\box0}$}   \def\ub{\underbar}    \def\ul{\underline}
\def\f{\left} \def\g{\right} \def\e{{\rm e}} \def\o{\over}
\def\vf{\varphi} \def\pl{\partial} \def\cov{{\rm cov
}} \def\ch{{\rm ch}}
\def\la{\langle} \def\ra{\rangle} \def\EE{e$^+$e$^-$}
\def\bitz{\begin{itemize}} \def\eitz{\end{itemize}}
\def\btbl{\begin{tabular}} \def\etbl{\end{tabular}}
\def\btbb{\begin{tabbing}} \def\etbb{\end{tabbing}}
\def\beqar{\begin{eqnarray}} \def\eeqar{\end{eqnarray}}
\def\\{\hfill\break} \def\dit{\item{-}} \def\i{\item}
\def\bbb{\begin{thebibliography}{9} \itemsep=-1mm}
\def\ebb{\end{thebibliography}} \def\bb{\bibitem}
\def\bpic{\begin{picture}(260,240)} \def\epic{\end{picture}}
\def\akgt{\noindent{\bf Acknowledgements}}
\def\fgn{\noindent{\bf\large\bf Figure captions}}
\def\fsz{\footnotesize}
\def\ifmath#1{\relax\ifmmode #1\else $#1$\fi}%
\def\rmt{\ifmath{{\mathrm{t}}}} \def\rmcut{\ifmath{{\mathrm{cut}}}}
\newcommand{\QCD}{{\sc qcd}} \newcommand{\NFM}{{\sc nfm}}
\newcommand{\BNL}{{\sc bnl}} \newcommand{\RHIC}{{\sc rhic}}
\newcommand{\CERN}{{\sc cern}} \newcommand{\LHC}{{\sc lhc}}
\newcommand{\ALICE}{{\sc alice}}
\def\pt{{p_{\rmt}}} \def\vf{\varphi} \def\yct{y_{\rmcut}} \def\kt{k_{\rmt}}
\def\levy{L$\acute{\rm e}$vy} \def\renyi{R$\acute{\rm e}$nyi}

\begin{document}


\footnotetext[0]{chengang1@cug.edu.cn}

\cl{\Large Study on the method of mini-jet identification}
\vskip0.3cm

\cl{\Large  in relativistic heavy ion collisions\footnote{ Supported
by National Natural Science Foundation of China (10775056).  }}

\vskip1.0cm \cl{\large \  Li De-sheng$^{(1)}$ \quad Tian
Feng-Ge$^{(1)}$ \quad Chen \ \ Gang$^{(1,2)}$ }

\vskip0.3cm \cl{\small $^{(1)}$ School of Mathematics and Physics,
China University of Geosciences, Wuhan, China, 430074}
 \cl{\small
$^{(2)}$ Key Laboratory of Quark \& Lepton Physics (CCNU), Ministry
of Education,China,430079}

\footnotetext[2] {Have submitted to the Chinese Physics C}

\date{ }

\begin{abstract}
 In this thesis a set of methods identifying minijet from
final state particles in the relativistic heavy ion collision events
is established and the parameter dependence has been investigated in
Au+Au collisions at $\sqrt s= 200$GeV using  a multiphase transport
model (AMPT). It is found that the number of minijets reduces with
the increasing of collision parameter and raises with the increasing
of c.m energy.
 Furthermore, we analyze the rapidity and momentum distribution inside
  minijets identified using this method.
\end{abstract}




\section{Introduction}
The basic theory of strong interactions---Quantum Chromodynamics
(QCD), has two distinct characters: asymptotic free and color
confinement. Thus, partons produced in high energy collisions will
turn to final flavorless hadrons before being observed. When the
virtuality $Q$ of a parton is high enough, the final hadrons will
form a taper structure around the direction of the origin parton,
which is called a jet. Jets, being experimentally observable, are
taken as an efficient way to study the physical properties and
interaction dynamics of partons.

In 1975, the two-jets structure was observed in e$^+$e$-$ collision
experiments at $\sqrt s \leq 6$ GeV~{\cite{cite1}}. This is regarded
as an adequate experimental evidence for the existing of partons,
which is predicted by the parton models {\cite{cite2}}.
Theoretically, as the energy increases, the quark -antiquark pairs,
moving in the opposite direction may emit a hard gluon with large
transverse momentum to form the third jet. In 1979, this astonishing
prediction of QCD was confirmed by experiments, when a third jet was
observed in e$^+$e$^-$ collisions at $\sqrt s = 17$ -- 30 GeV, which
is considered as the earliest experimental evidence for gluon's
existence{\cite{cite3,cite4,cite5,cite6}}.

In nucleus-nucleus and hadron-hadron collisions, due to the
existence of a large background the situation is more complicated.
For the basic theory --- QCD remains valid here, jets' production is
still anticipated when the collision energy is high enough. In early
the 1990s, the production of jets in hadron-hadron collisions was
widely studied {\cite{cite7,cite8,cite9,cite10,cite11}} and had been
considered as an efficient way to obtain the strong coupling
constant $\alpha_{\rm _S}${\cite{cite12,cite13}}.

In the last century, nucleus-nucleus collisions were carried out at
CERN. However, because the center of mass energy $\sqrt{s_{NN}}$ in
those fixed target experiment was less than 20GeV, jets predicted
beforehand were not observed. In the beginning of this century, the
first relativistic heavy ion collider RHIC at BNL successfully
realized Au-Au collision at $\sqrt{s_{NN}}$ =200GeV, jet production
becames available. In the collisions, the jets were observed to have
energy loss when they crossed the central area
{\cite{cite14,cite15,cite16,cite17}}, which is called jet quenching.
The observation of this phenomenon is regarded as one of the main
accomplishments in RHIC experiments; and also it is taken as the
implication for the configuration of heat and high density matters
produced in relativistic heavy ion collisions{\cite{cite18,cite19}}.
Thus, jet is recognized as a powerful tool for studying the
properties of the produced new form of matter{\cite{cite20,cite21}}.

Due to the significance of jet physics, in this thesis, we carefully
study the methods for selecting minijets and the characters of
selected minijets in relativistic heavy ion collisions.

\section{The method for selecting minjets}
\label{sec} Theoretically{\cite{cite22}}, jet is a cone structure
formed by a group of hadrons or partons around one certain axis. In
high energy hadron-hadron collisions, jet is generally defined with
the cone algorithm{\cite{cite23,cite24}}. One certain direction is
chosen as the jet axis and then the pseudo-rapidity $\eta$ and
azimuthal angle $\varphi$ plane for a particle are constructed,
where $\eta$ is defined as $\eta=-\frac{1}{2}\ln \tan \theta$, and
$\theta$ is the angle with jet axis. With defining a radius
parameter $R=\sqrt{\eta^2+\varphi^2}$ for each particle, the
particles with $R\leq R_0$ are identified as a
jet{\cite{cite25}},where

\begin{equation}\label{eq:}
\varphi=\arccos{\frac{p_x}{p_T}}.
\end{equation}

In $e^+e^-$ collisions, jets are generally identified through some
jet-finding processes, e.g the Jade or Durham {\cite{cite26}}, in
which there is a parameter $y_{cut}$. This relationship between
$y_{cut}$ and the cut value of relative transverse momentum
$k_{t,cut}$ {\cite{cite27}} is
\begin{equation}\label{eq2}
k_{t,cut}=\sqrt{y_{cut}}\cdot\sqrt{s},
\end{equation}
where $\sqrt{s}$ is the c.m energy. The relative transverse momentum
$k_t$ between two particles $i$ and $j$ is defined as
{\cite{cite27}}
\begin{equation}\label{eq2}
k_{t,ij}=2\min{(E_i,E_j)}\sin{\frac{\theta_{ij}}{2}}.
\end{equation}

Two particles having a $k_t$ smaller than $k_{t,cut}$ are grouped
into one jet\cite{cite27}.

In relativistic heavy ion collisions, people usually use a high
transverse momentum trigger to define a jet, and the chosen trigger
is called a leading particle. Assume the momentum, transverse
momentum, rapidity and azimuth angle of the leading particle are
$P,p_t$, $y$ and $\varphi$, and take this particle as the jet
center; and the momentum, transverse momentum, rapidity and azimuth
angle of the around particle $i$ are denoted by $p_i$, $p_{ti}$,
$y_i$and $\varphi_i$, respectively. Jet is generally considered as a
cone structure formed by a group of particles, so the rapidity and
azimuth angles, relative to the leading particle, of all particles
inside one jet should fall into a certain range. Thus we define the
distance between the leading particle and particle $i$ as follows:
\begin{equation}\label{eq2}
R_i=\sqrt{(y-y_i)^2+(\varphi-\varphi_i)^2}.
\end{equation}
If we choose a radium cut parameter $R_0$ to define a cone space,
when the distance of $i$th particle with the leading particle $R_i <
R_0$, this particle will fall into this cone space.

In summary, in relativistic heavy ion collisions, the minijets may
be selected as the following steps:
\begin{enumerate}

\item Firstly, the parameters for selecting minijets, the maximum radius $R_0$
and the leading particles' minimum transverse momentum $P_{T0}$, are
fixed on.

\item A particle with the largest transverse momentum is selected out
from all final particles, and is considered as the leading
particles. The transverse momentum, rapidity and azimuth angle of
leading particle are labeled as $p_t$, $y$, $\varphi$, respectively.
This leading particle is taken as the center to select the particles
forming a minijet.


\item According to Equation (4),  the distance $R_i$ between  the leading particle and
the other particle $i$ is calculated. If $R_i<R_0$,  the particle
$i$ belongs to this minijet.
\end{enumerate}

 The above steps in the remainder particles are repeated until there is
 no particle with transverse momentum large enough to be selected as a leading particle.

As an example, a sample event of relativistic heavy ion Au+Au
collision at $\sqrt{s_{NN}}=200GeV$ is produced by  using a
multiphase transition model (AMPT){\cite{cite28}}, and the minijets
are identified using the process mentioned above.

First of all, to analyze the influence of maximum radius, we select
minijets when radius $R_0$ are $0.5,0.75, 1.00,1.25, 1.5,1.75$ and
$2.0$,and the transverse momentum $P_{T0}$ of the leading particle
varies from $0.5$ to 2.9 GeV, respectively. The results for the
numbers of selected out minijets are shown in Fig.1.

As shown in Fig.1, the cut parameters significantly influence the
selection of minijets. Obviously, if the transverse momentum
$P_{T0}$ of the leading particle selected minijet  are fixed, the
total number of minijet reduces as the cut parameter $R_{0}$
increases. Because the sum of particles is definite, with larger
$R_0$, one jet contains more particles, as a consequence the number
of jets reduces. On the other hand with larger minimum transverse
momentum set for leading particles, the number of particles fitting
the criteria reduces, and it is the same case with the number of
minijets. Obviously, in high energy heavy ion collisions, the
minijet selection has the dependence on cut parameters. However,
when $R_{0}\ge 1.5,P_{T0}\ge 2.0$GeV, this dependence on cut
parameters $R_{0}$ and $P_{T0}$ will be weakened.

\begin{center}
\includegraphics[width=8.cm]{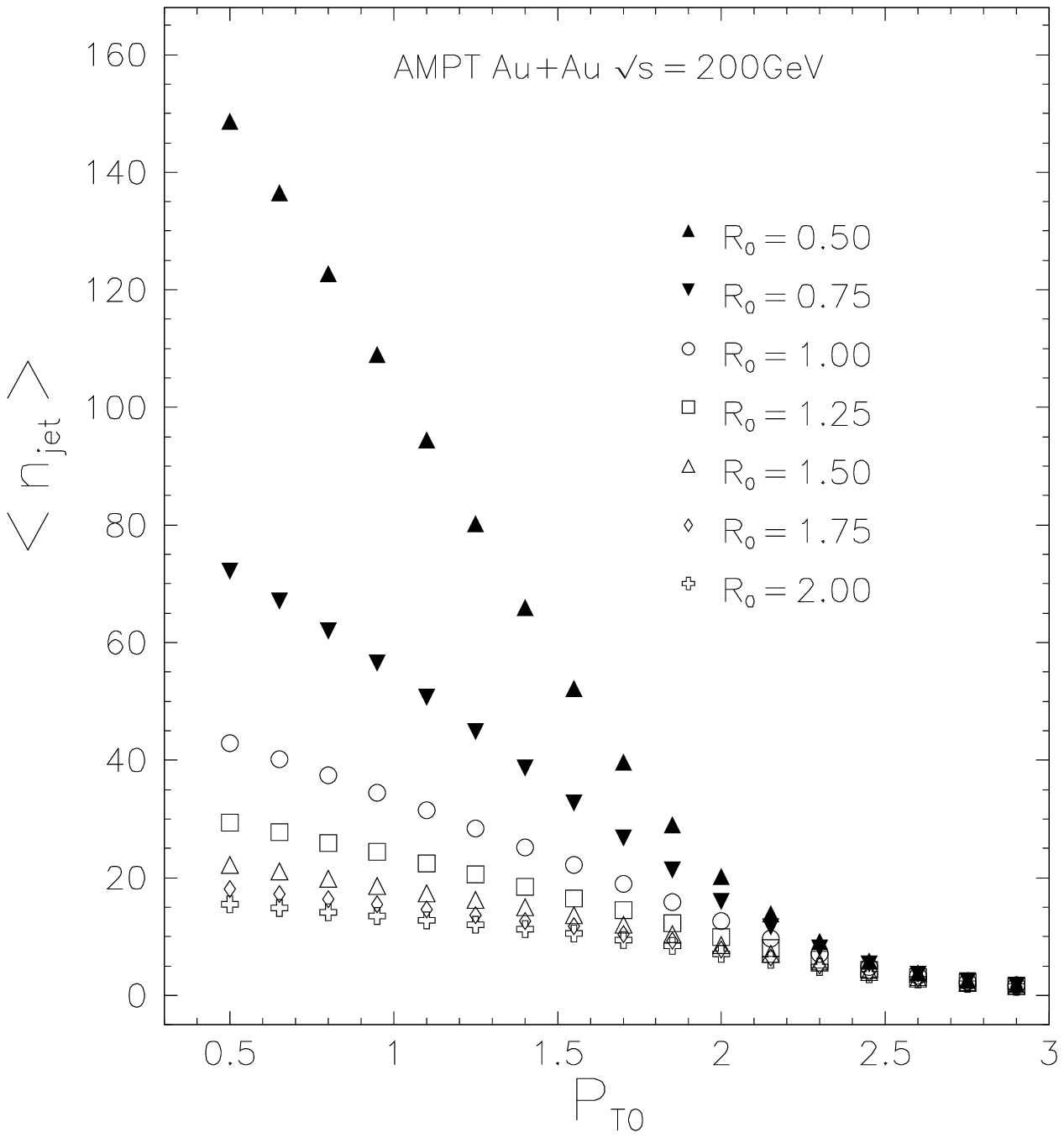}

{fig1\  The variation of the average number of minijets $\langle
n_{jet}\rangle $ with the transverse momentum $P_{T0}$ of leading
particle for AMPT Au-Au collision at $\sqrt s=200$GeV, and the
dependence of selected minijets numbers on cut parameters $R_0$.}
\end{center}

\section{ The dependence of minijets on energy and collision parameters }

\label{sec} Now we apply the above-mentioned minijet-algorithm to
analyze the event samples of the relativistic heavy ion Au+Au
collision with different values of c. m. energy and collisions
parameter $b$, and study their dependence of the total number of
minijet selected from one event on the different value of c.m.
energy and collision parameter $b$. The event samples of different
c.m. energy and collision parameter $b$ are constructed from AMPT,
respectively, each consists of 1000 relativistic heavy ion Au+Au
collision events.

Under the condition for the selection parameters as
$R_0=1.5,P_{T0}=2$GeV and $b=0$, we calculate the average number of
minijet selected when c.m. energy equals to $20,40,\cdots,200$ GeV,
respectively. The result is shown in Fig.2(a).

In the same way, under the condition for the selection parameters as
$R_0=1.5,p_{t0}=2$GeV and $\sqrt s=200$ GeV, we calculate the
average number of minijet selected when collision parameter $b$
equals  $1,2,\cdots,8$ fm,respectively. The result is shown in
Fig.2\ (b).

 Fig.2\ (a) shows that the number of minijets raises with the c.m. energy increasing,
 which is because that the
higher the collision energy is, the more fierce the collision is,
i.e, the number of total particles and leading particles raises. So
the number of minijets raises with the c.m. energy increasing.

\begin{center}
\includegraphics[width=12cm]{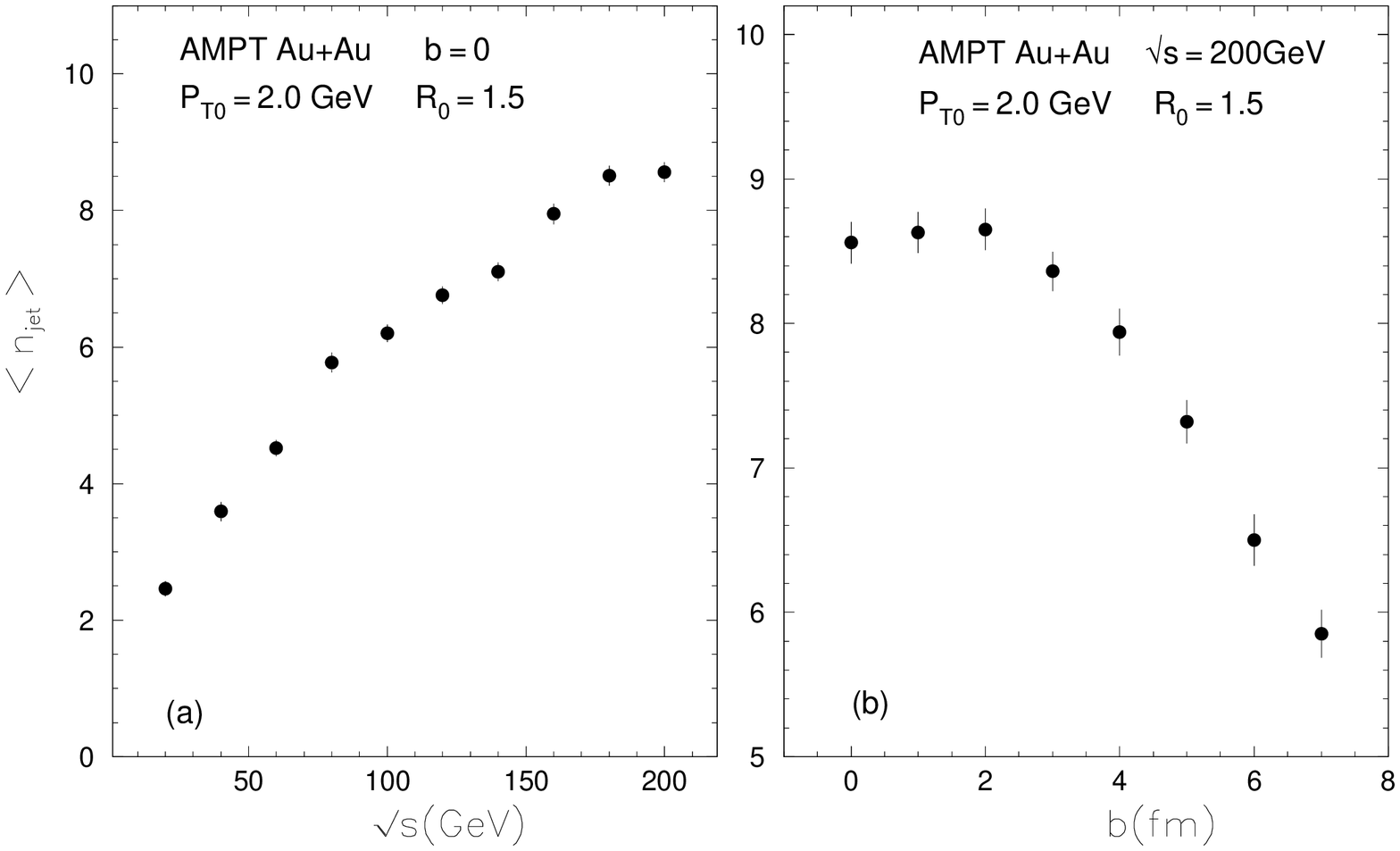}

{Fig2\  The distribution of average number of minijets (a) as a
function of the collision energy $\sqrt s$  at the collision
parameter $b=0$, (b) as a function of the collision parameter $b$ at
$\sqrt s=200$GeV, the data of Au-Au collisions produced by AMPT
model.}
\end{center}

It can be seen from  Fig.2\ (b) that the number of minijets reduces
with the increasing of collision parameter. This is because that the
collision becomes weak when the collision parameter $b$ increases,
so the number of total particles and leading particles reduces. Thus
the number of minijets reduces as the collision parameter increases.

\section{ The longitudinal and transverse distributions
of particles inside minijets }

\label{sec} In order to study the particle distributions inside the
minijets, a full event samples of final state particles with 1000
relativistic heavy ion Au+Au collision events at c.m. energy
$\sqrt{s_{NN}}=200$ GeV are generated using AMPT {\cite{cite28}.
Then the events subsamples of minijets are determined, using
minijet-algorithm given previously selected from the full event
sample, with the cut parameters selecting minijets being chosen as
$R_0=1.5, b=0$, respectively. Then we analyze the variance of
rapidity and transverse momentum inside minijets.

Firstly, the momentums of all the particles in the minijet are
summed up to $p_{jet}$, which is defined as the jet momentum. The
direction of $p_{jet}$ is defined as the longitudinal direction and
the directions perpendicular to it are the transverse directions.
The rapidity $y$ and transverse momentum $p_t$ inside minijet are
defined according to these directions as usual.
\begin{center}
\includegraphics[width=10cm]{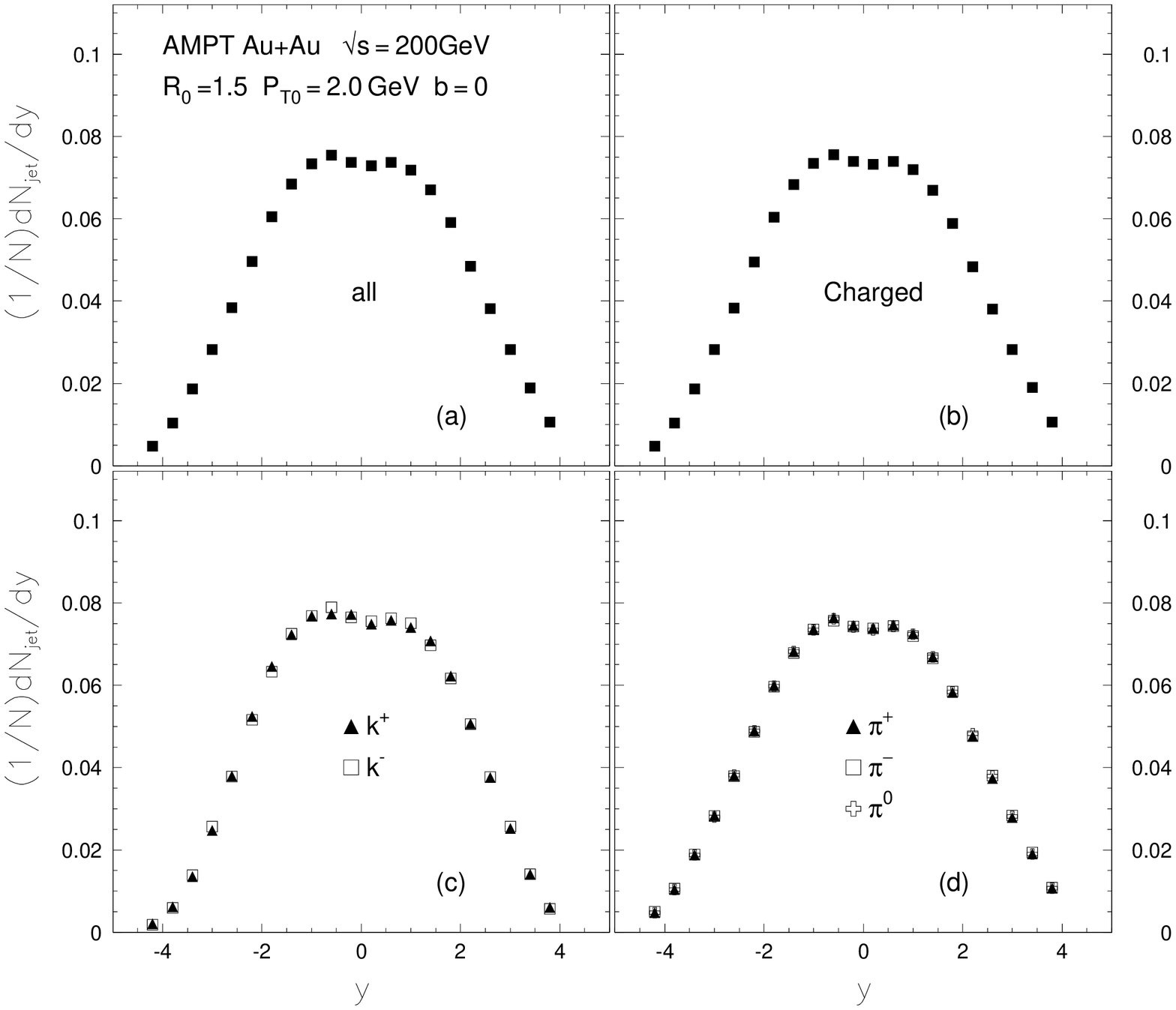}

{Fig3\ Rapidity distributions of minijets from Au+Au collision at
$\sqrt s=200$GeV, with cut parameter $R_0=1.5$, and collision
parameter $b=0$, (a) all particles, (b) charged particles, (c)
$K^+,K^-$, (d) $\pi^+,\pi^-,\pi^0$. }
\end{center}

The rapidity distributions for all particles, charged particles,
 Kaons and Pions inside a single minijets are shown in Fig.3,
respectively. It can be seen from Fig.3(a) that the rapidity
distributions are distributed over the -4.2 to 4.2 district and most
of the particles inside minijets are distributed in the center
rapidity range. The number of particles increases sharply from $y =
-4$ to $-1$, then turns to be an invariant flat until $y=1$ and
quickly decreases as the rapidity is more than 1. The rapidity
distributions inside a single minijet for charged particles, Kaons
($K^+,K^-$) and Pions ($\pi^{+},\pi^{-},\pi^{0}$) shown in Figure
3(b)-(c), have the similar behavior.

Fig.4 shows the transverse momentum distributions of particles
inside a single minijets corresponding to all particles, charged
particles, Kaons and Pions, respectively. The transverse momentum
distributions increase sharply from $p_t = 0$ until a maximum peak
around $p_t = 0.2 \sim 0.26$ is reached. The transverse momentum
distributions of different particle inside a single minijet, shown
in Fig.4(b)-(d), have a similar behavior, which the peak moves
rightward with the increasing of mass of particle comparison
Fig.4(c). Contrary to the dependence of transverse momentum
distributions on the mass of the particle inside a single minijet,
it turns out to be independent on their charge (in Fig.4(c) or (d)).

\begin{center}
\includegraphics[width=10cm]{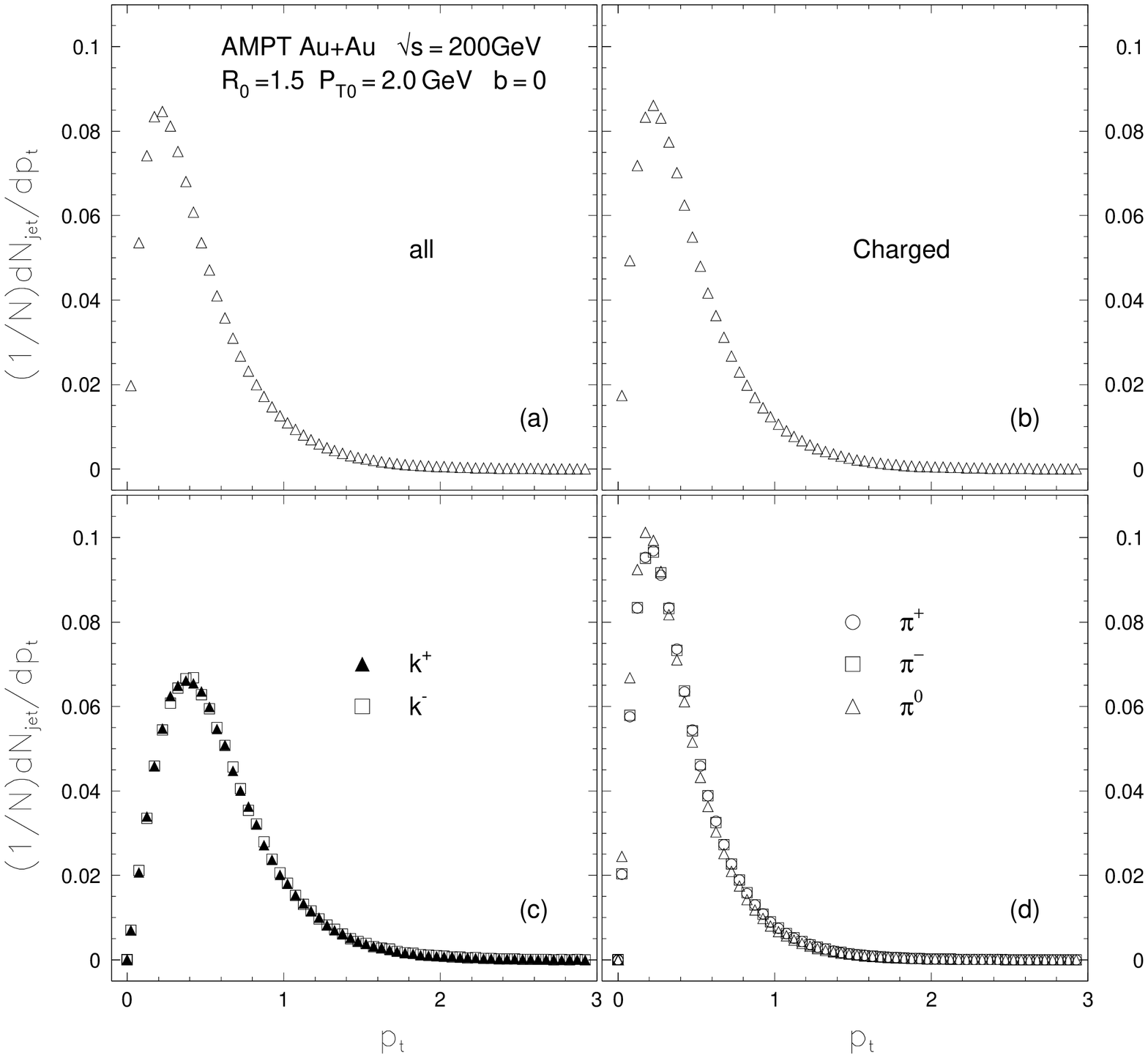}

{Fig4\ Transverse momentum distributions of minijets from Au+Au
collision at $\sqrt s=200$GeV, with cut parameter $R_0=1.5 $,and
collision parameter $b=0$, (a) all particles, (b) charged particles,
(c) $K^+,K^-$, (d) $\pi^+,\pi^-,\pi^0$.}
\end{center}
\section{Conclusion and discussion}

In this paper we have defined a parameters ---- the relative
distance cut parameter $R_0$, and a set of methods identifying
minijets from final state particles in the relativistic heavy ion
collision events is carried out. The dependence of this method
identifying minijets on
  cut parameters is studied  in relativistic heavy ion Au+Au collision
  events at $\sqrt s= 200$GeV using  AMPT. Obviously,
  in high energy heavy ion collisions, the jet selection
has the dependence on cut parameters. As $R_0 \ge 1.5$ and
$P_{T0}\ge 2.0$GeV, this dependence on cut parameters will be
weakened. On the other hand, it is found that the number of minijets
selected using this method reduces with the increasing of collision
parameter, and raises when the c.m. energy increases.
 Furthermore, we have analyzed the distribution properties of rapidity  and momentum inside
minijets in high energy heavy ion collisions.

It is worth noting that an event consists of large hard scattering
and also many soft collisions in the relativistic heavy ion
collision. The minijet should originate in the hard scattering.
However, we have selected minijet which consists of large transverse
momentum outgoing hadrons that originate from the large transverse
momentum partons and also hadrons that originate from the ''soft''
or ''semi-hard'' multiple parton interactions. Since the existence
of such soft interactions, it makes the number of selecting minijet
errors and the particles inside minijet not to be pure. The
distribution characteristics of particles inside minijet have
certain uncertainty.


\vspace{-1mm} \centerline{\rule{80mm}{0.1pt}} \vspace{2mm}



\clearpage

\end{document}